\documentclass[showpacs,preprint,aps]{revtex4}
\usepackage{amsmath}
\usepackage{dcolumn}
\usepackage{bm}
\usepackage{graphicx}
\usepackage{amsfonts}
\usepackage{amssymb}

\begin{document}

\title{Comparison of different measures for quantum discord under non-Markovian noise}
\author{Z. Y. Xu$^{1,2}$}
\email{zhenyuxu.wipm@gmail.com}
\author{W. L. Yang$^{1}$}
\author{X. Xiao$^{3}$}
\author{M. Feng$^{1}$}
\email{mangfeng@wipm.ac.cn}
\affiliation{$^{1}$State Key Laboratory of Magnetic Resonance and Atomic and Molecular
Physics, Wuhan Institute of Physics and Mathematics, Chinese Academy of
Sciences, and Wuhan National Laboratory for Optoelectronics, Wuhan 430071, China\\
$^{2}$Graduate School of the Chinese Academy of Sciences, Beijing 100049,
China\\
$^{3}$Key Laboratory of Low-Dimensional Quantum Structures and Quantum
Control of Ministry of Education, Department of Physics, Hunan Normal
University, Changsha 410081, China}

\begin{abstract}
Two geometric measures for quantum discord were recently proposed by
Modi \textit{et al}. [Phys. Rev. Lett. \textbf{104}, 080501 (2010)]
and Daki\'{c} \textit{et al}. [Phys. Rev. Lett. \textbf{105}, 190502
(2010)]. We study the similarities and differences for total quantum
correlations of Bell-diagonal states using these two geometry-based
quantum discord and the original quantum discord. We show that,
under non-Markovian dephasing channels, quantum discord and one of
the geometric measures stay constant for a finite amount of time,
but not the other geometric measure. However, all the three measures
share a common sudden change point. Our study on critical point of
sudden transition might be useful for keeping long time total quantum
correlations under decoherence.
\end{abstract}

\maketitle

Entangled states cannot be prepared by local operations and classical
communication \cite{entanglement,book QIC}. One may think that
the exchange of classical information would not add any quantum correlation
to the state. This is true for pure states but not for a general mixed
state, because quantum correlations also exist in some mixed separable
states and have played important roles in some quantum tasks, such as in
deterministic quantum computation with one pure qubit \cite{DQC1}. To
capture the total quantum correlations, a measure called quantum discord has
been first proposed by Olliver and Zurek \cite{discord} and by Henderson and
Vedral \cite{discord-dengjia}, and then widely studied \cite%
{add1,D-Maxwell,add2,D-2qubit,D-XX,NLB,D0,D-fields,Hundun,D-X,D-NMR,D-CV,MD,
SCDchan,SCD-exp,SCDcons1,SCDcons2,SCDchan2,nonMD,discord-R,discord-G,discord-GL}.

Quantum discord is spoiled due to unavoidable interaction between the
quantum system and the surrounding environment. The dynamics of quantum
discord has been investigated under both Markovian \cite%
{MD,SCDchan,SCDcons1,SCDcons2,SCD-exp,SCDchan2} and non-Markovian \cite%
{nonMD} environments. Of particular interest is that, for a special
class of quantum Bell-diagonal states, there exists sudden change of
quantum discord under Markovian environment \cite{SCDchan}.
Moreover, the constant quantum discord under Markovian phase-damping
channels was observed experimentally \cite{SCD-exp} and intensively
studied theoretically \cite{SCDcons1,SCDcons2}.

Quantum discord (capturing total quantum correlations) has also been explored
from the aspect of geometry, where two measures were recently
proposed based on, respectively, the relative entropy
\cite{discord-R} and the square of Hilbert-Schmidt norm
\cite{discord-G}. The present work concentrates on the comparison of these
two geometric measures with the originally defined quantum discord.
Since some interesting features, such as suddenly changing and constant
total quantum correlations, have been discovered by the quantum
discord with respect to Bell-diagonal states, we wonder if these
features remain in the two geometric measures. We will focus on
non-Markovian environments, from which we could obtain some insights
into the protection of total quantum correlations from decoherence.

Quantum discord is defined as a measure of the discrepancy between two
different quantum analogs of the classical mutual information \cite%
{discord,discord-dengjia}. For a bipartite system $\rho _{AB}$, the quantum
discord is given by
\begin{equation}
\mathcal{D}\left( \rho _{AB}\right) =\mathcal{I}\left( \rho _{AB}\right) -%
\mathcal{C}\left( \rho _{AB}\right) ,
\end{equation}%
where $\mathcal{I}\left( \rho _{AB}\right) =S\left( \rho _{A}\right)
+S\left( \rho _{B}\right) -S\left( \rho _{AB}\right) $ is the
quantum mutual information (also called total correlations \cite
{discord-dengjia}), with $S\left( \rho \right) =-$Tr$(\rho \log
_{2}\rho )$ the von Neumann entropy of $\rho $ and $\rho _{A(B)}$
the reduced density matrix of $\rho _{AB}$ by tracing out $B(A)$.
$\mathcal{C}\left( \rho _{AB}\right) =\max\limits_{B_{k}}\{S\left(
\rho _{A}\right) -\sum_{k}q_{k}S\left( \rho _{A}^{k}\right) \}$ is
considered as the classical correlation, where $\rho
_{A}^{k}=$Tr$_{B}\left\{ B_{k}\rho _{AB}B_{k}^{\dag }\right\}
/q_{k}$ is the resulting state after the measurement $\left\{
B_{k}\right\} $ on $B$, and $ q_{k}=$Tr$_{AB}\left\{ B_{k}\rho
_{AB}B_{k}^{\dag }\right\}.$ Note that quantum discord is not
symmetric with respect to exchanging A and B, however, for the
Bell-diagonal states under consideration, it is\cite{discord-G}.

From the relative entropy perspective, the geometric measure $Q_{\mathcal{R}}$
quantifies how distinguishable a given state $\rho $ is from the closest
classical state $\upsilon $ \cite{discord-R},
\begin{equation}
Q_{\mathcal{R}}\left( \rho \right) =\min_{\upsilon \in \mathcal{G}}S\left(
\rho ||\upsilon \right) ,
\end{equation}%
where $\mathcal{G}$ is the set of classical states \cite{Note0} and $S\left( \rho
||\upsilon \right) =-$Tr$\left( \rho \log _{2}\upsilon \right) -S\left( \rho
\right) $ is the relative entropy.

Another geometric measure $Q_{\mathcal{S}}$ is defined based on the fact
that almost all quantum states have non-vanishing quantum discord \cite%
{D0,discord-G}. The distance between a given state $\rho $ and the nearest
zero-discord state $\varrho $ is defined as \cite{discord-G},
\begin{equation}
Q_{\mathcal{S}}\left( \rho \right) =\min_{\varrho \in \Omega }||\rho
-\varrho ||^{2},
\end{equation}
where $\Omega $ denotes the set of quantum states with zero-discord
\cite{Note0} and $||\rho -\varrho ||^{2}=$Tr$(\rho -\varrho )^{2}$
is the square of Hilbert-Schmidt norm of Hermitian operators
\cite{discord-G,discord-GL}. For a two-qubit system $\rho
_{AB}=(\mathbf{1\otimes 1}+\sum_{j=1}^{3}\alpha _{j}\sigma
_{j}^{A}\otimes \mathbf{1}+\sum_{j=1}^{3}\beta _{j}\mathbf{1}
\otimes \sigma _{j}^{B}+\sum_{j,k=1}^{3}\mathcal{M}_{jk}\sigma
_{j}^{A}\otimes \sigma _{k}^{B})/4$, with $\mathbf{1}$ and \{$\sigma
_{j}$\}
being the identity and Pauli operators, Eq. (3) can be simplified as%
\begin{equation}
Q_{\mathcal{S}}\left( \rho _{AB}\right) =\frac{1}{4}\left( ||\vec{\alpha}%
||^{2}+||\mathcal{M}||^{2}-\delta _{\max }\right),
\end{equation}
where $\vec{\alpha}=\left( \alpha _{1},\alpha _{2},\alpha
_{3}\right) ^{T}$ is a column vector,
$\mathcal{M}=(\mathcal{M}_{jk})$ is a matrix, and $ \delta _{\max }$
is the largest eigenvalue of matrix $\vec{\alpha}\vec{\alpha
}^{T}+\mathcal{MM}^{T}.$ Here the superscript $T$ represents the
transpose of vectors or matrices.

We start our analysis by considering two identical qubits A and B initially
in a Bell-diagonal state with the density operator as
\begin{equation}
\rho _{AB}(0)=\frac{1}{4}\left( 1+\sum_{j=1}^{3}c_{j}\sigma _{j}^{A}\otimes
\sigma _{j}^{B}\right) =\sum_{a,b=0,1}\lambda _{ab}|\chi _{ab}\rangle
\left\langle \chi _{ab}\right\vert ,
\end{equation}
where the eigenstates are four Bell states $|\chi _{ab}\rangle =\textsf{%
\textbf{(}}|0,b\rangle +(-1)^{a}|1,1\oplus b\rangle \textsf{\textbf{)}}/%
\sqrt{2}$ with eigenvalues $\lambda _{ab}=\textsf{\textbf{(}}%
1+(-1)^{a}c_{1}-(-1)^{a+b}c_{2}+(-1)^{b}c_{3}\textsf{\textbf{)}}/4$ (a,b=0,1)\cite%
{D-2qubit}$,$ and $(c_{1},c_{2},c_{3})$ are three parameters of the
Bell-diagonal states. Considering $\lambda _{ab}\geq 0,$ all Bell-diagonal
states should be confined within a tetrahedron in three-dimensional space
spanned by $c_{1}$, $c_{2}$, and $c_{3}$ \cite{simianti}.

In what follows, we consider the situation of the qubits under independent
non-Markovian dephasing channels \cite{nonM}. The dynamics of the qubits can
be characterized by the Kraus operators \{$K_{\mu }(t)$\}: $\rho
_{AB}(t)=\sum_{\mu }K_{\mu }(t)\rho _{AB}(0)K_{\mu }^{\dag }(t),$ where the
Kraus operators satisfy $\sum_{\mu }K_{\mu }^{\dag }(t)K_{\mu }(t)=1.$ The
Kraus operators for this non-Markovian model are given by $K_{\mu
}(t)=\kappa _{a}(t)\otimes \kappa _{b}(t)$ $(a,b=0,1)$ where $\kappa
_{0}(t)=\left(
\begin{array}{cc}
\omega (t) & 0 \\
0 & 1
\end{array}
\right) $ and $\kappa _{1}(t)=\left(
\begin{array}{cc}
\sqrt{1-\omega ^{2}(t)} & 0 \\
0 & 0
\end{array}
\right) $, with $\omega (t)=\exp \textsf{\textbf{(}}-f(t)\textsf{\textbf{)}}$%
, $f(t)=\Gamma \textsf{\textbf{(}}t+(e^{-\gamma t}-1)/\gamma \textsf{\textbf{%
)}}/2$, $\gamma $ denotes the environmental noise bandwidth and $\Gamma $ is
the Markovian decay rate. Explicitly, the time evolution of the system can
be expressed as $\rho _{AB}(t)=\sum_{a,b=0,1}\lambda _{ab}(t)|\chi
_{ab}\rangle \left\langle \chi _{ab}\right\vert ,$ where $\lambda _{ab}(t)=
\textsf{\textbf{(}}1+(-1)^{a}c_{1}(t)-(-1)^{a+b}c_{2}(t)+(-1)^{b}c_{3}(t)
\textsf{\textbf{)}}/4,$ $c_{1}(t)=c_{1}(0)\omega ^{2}(t),$ $
c_{2}(t)=c_{2}(0)\omega ^{2}(t)$, and $c_{3}(t)=c_3$ [$c_{3}(t)$ is constant during the evolution].
For simplicity, we
denote in the following by $c_{2}(t)=\epsilon c_{1}(t),$ with $\epsilon
=c_{2}(0)/c_{1}(0)$. In addition, the above results could return to
the Markovian situation by setting $f(t)\rightarrow \Gamma t/2$ in the
Markovian limit $\gamma \rightarrow \infty$.

According to Ref. \cite{D-2qubit}, the classical correlation is calculated
as
\begin{eqnarray}
\mathcal{C}\textsf{\textbf{(}}\rho _{AB}(t)\textsf{\textbf{)}} &\mathbf{=}%
&\sum_{l=0,1}\frac{1+(-1)^{l}\mathbf{m}}{2}\log _{2}\textsf{\textbf{(}}%
1+(-1)^{l}\mathbf{m}\text{$\textsf{\textbf{)}}$}\mathbf{,}  \notag \\
&=&1-H_{bin}\left( \frac{1+\mathbf{m}}{2}\right) ,
\end{eqnarray}%
where $\mathbf{m}=\max \{|c_{1}(t)|,|c_{2}(t)|,|c_3|\}$ and $H_{bin}\left(
p\right) =-p\log _{2}p-(1-p)\log _{2}(1-p)$ is the binary entropy \cite{book
QIC}. In addition, the total correlation is $\mathcal{I}\textsf{\textbf{(}}%
\rho _{AB}(t)\textsf{\textbf{)}}$$\mathbf{=}2\mathbf{+}\sum_{a,b=0,1}\lambda
_{ab}(t)\log _{2}\lambda _{ab}(t)$ \cite{D-2qubit}, which, for the initial
conditions $|c_{1}(0)|\geq |c_{2}(0)|,|c_3|$ and $\epsilon =-c_3$ \cite{Note},
can be expressed as
\begin{eqnarray}
\mathcal{I}\textsf{\textbf{(}}\rho _{AB}(t)\text{$\textsf{\textbf{)}}$} &%
\mathbf{=}&\sum_{\substack{ l=0,1  \\ x=c_3,c_{1}(t)}}\frac{1+(-1)^{l}x}{2}%
\log _{2}\text{$\textsf{\textbf{(}}$}1+(-1)^{l}x\text{$\textsf{\textbf{)}}$},
\notag \\
&=&2-H_{bin}\left( \frac{1+c_{1}(t)}{2}\right) -H_{bin}\left( \frac{1+c_3}{2}%
\right) .
\end{eqnarray}
Therefore, according to Eq. (1), the quantum discord is given by
\begin{equation}
\mathcal{D}\text{$\textsf{\textbf{(}}$}\rho _{AB}(t)\text{$\textsf{\textbf{)}%
}$}\mathbf{=}\left\{
\begin{array}{cc}
1\mathbf{-}H_{bin}\left( \frac{1+c_3}{2}\right) \text{\textbf{,}} & t\leq \tau
, \\
1\mathbf{-}H_{bin}\left( \frac{1+c_{1}(t)}{2}\right) \text{\textbf{,}} &
t>\tau ,%
\end{array}%
\right.
\end{equation}%
where
\begin{equation}
\tau =\frac{1+\eta \gamma +\mathcal{W}(-e^{-1-\eta \gamma })}{\gamma }
\end{equation}%
is the critical point with $\eta =-\frac{\ln \left\vert
\frac{c_3}{c_{1}(0)} \right\vert }{\Gamma }$ and $\mathcal{W}(\cdot
)$ the Lambert $\mathcal{W}$ function. This is really an interesting
phenomenon, since it seems to exist a `decoherence-free' area of
total quantum correlations when $t\leq \tau $ \cite
{SCD-exp,SCDcons1,SCDcons2} [shown in Fig. 1(a)].

%%%%%%%%%%%%%%%%%%%%%%%%%%%%%%%%%%%%%%%%%
\begin{figure}[tbp]
\centering
\includegraphics[width=4.0in]{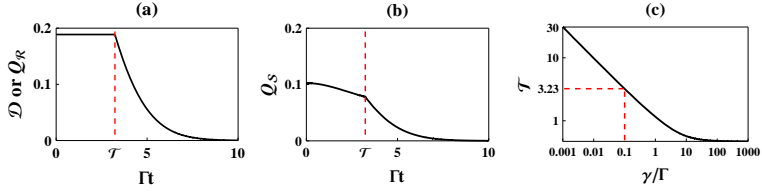}
\caption{Dynamics of total quantum correlations using $\mathcal{D}$ or $Q_{\mathcal{R}}$
in (a) and using $Q_{\mathcal{S}}$ in (b) as a function of $\Gamma t
$ for Bell-diagonal state with the initial conditions \textsf{\textbf{(}}$%
c_{1}(0),c_{2}(0),c_{3}(0)$\textsf{\textbf{)}}$\mathbf{=(}0.8,-0.4,0.5%
\textbf{)}$ under non-Markovian dephasing channels with $\protect\gamma %
/\Gamma =0.1.$ (c) is for the critical point $\mathcal{T}=\protect\tau \Gamma $
with $\left\vert c_3/c_{1}(0)\right\vert =5/8$ as a function of dimensionless
scaled reservoir bandwidth $\protect\gamma /\Gamma $ ranging from
non-Markovian regime to Markovian regime, where $\Gamma $ is the Markovian
decay rate.}
\end{figure}
%%%%%%%%%%%%%%%%%%%%%%%%%%%%%%%%%%%%%%%%%

In Fig. 1(c), we have plotted the critical point $\mathcal{T}(=\Gamma \tau )$
as a function of dimensionless scaled reservoir noise bandwidth $\gamma
/\Gamma .$ We found that the critical point $\mathcal{T}$ grows with the
decrease of the reservoir bandwidth. This implies that the non-Markovian
behavior would prolong the quantum correlation under decoherence. When $
\gamma \rightarrow \infty $ (Markovian limit)$,$ $\tau $ reduces to the
cases studied in Ref. \cite{SCDcons1}.

To calculate $Q_{\mathcal{R}}$, we denote the eigenvalues of
Bell-diagonal states in a decreasing order by $\lambda_{1}(t)\geq
\lambda_{2}(t)\geq \lambda_{3}(t)\geq \lambda_{4}(t)$. Therefore,
the closest classical states of $\rho _{AB}(t)$ are of the form
$\upsilon =\frac{\Lambda }{2}\textsf{\textbf{(}}|\lambda
_{1}(t)\rangle \left\langle \lambda _{1}(t)\right\vert +|\lambda
_{2}(t)\rangle \left\langle \lambda
_{2}(t)\right\vert\textsf{\textbf{)}}+\frac{1-\Lambda
}{2}\textsf{\textbf{(}}|\lambda_{3}(t)\rangle \left\langle
\lambda_{3}(t)\right\vert +|\lambda_{4}(t)\rangle \left\langle
\lambda_{4}(t)\right\vert\textsf{\textbf{)}}$ \cite{classical}, with
$\Lambda =\lambda _{1}(t)+\lambda _{2}(t).$ So the relative entropy
based quantum discord is given by
\begin{equation}
Q_{\mathcal{R}}\text{$\textsf{\textbf{(}}$}\rho
_{AB}(t)\text{$\textsf{\textbf{)}}$}\mathbf{=}\sum_{a,b=0,1}\lambda
_{ab}(t)\log _{2}\lambda _{ab}(t)+H_{bin}\left( \Lambda \right)
+1\text{\textbf{.}}
\end{equation}
We may find $H_{bin}\left( \Lambda \right) =H_{bin}\left(\frac{1+
\mathbf{m}}{2}\right)$ in both Markovian and non-Markovian regimes, which
implies that $Q_{\mathcal{R}}$ and $\mathcal{D}$ are equivalent for
Bell-diagonal states.

On the other hand, for the Bell-diagonal states under non-Markovian dephasing channels, the geometric measure of
quantum discord based on the square of Hilbert-Schmidt norm can be
obtained exactly as follows \cite{discord-G}
\begin{eqnarray}
Q_{\mathcal{S}}\text{$\textsf{\textbf{(}}$}\rho _{AB}(t)\text{$\textsf{%
\textbf{)}}$} &\mathbf{=}&\frac{1}{4}\text{$\textsf{\textbf{(}}$}%
c_{1}^{2}(t)+c_{2}^{2}(t)+c_{3}^{2}  \notag \\
&&-\max \{c_{1}^{2}(t),c_{2}^{2}(t),c_{3}^{2}\}\text{$\textsf{\textbf{)}}$%
}.
\end{eqnarray}
Clearly, with the initial conditions $|c_{1}(0)|\geq |c_{2}(0)|,|c_3|$ and $%
\epsilon =-c_3$, $c_{2}^{2}(t)$ will not be larger than $c_{1}^{2}(t)$ and $%
Q_{\mathcal{S}}$ is strongly dependent on the relation between $\left\vert
c_{1}(t)\right\vert $ and $\left\vert c_3\right\vert .$ Therefore, the geometric quantum
discord in such a case can be written as
\begin{equation}
Q_{\mathcal{S}}\text{$\textsf{\textbf{(}}$}\rho _{AB}(t)\text{$\textsf{%
\textbf{)}}$}\mathbf{=}\left\{
\begin{array}{cc}
\text{$\textsf{\textbf{(}}$}c_{2}^{2}(t)+c_{3}^{2}\text{$\textsf{\textbf{)}}$}/4%
\text{\textbf{,}} & t\leq \tau , \\
\text{$\textsf{\textbf{(}}$}c_{1}^{2}(t)+c_{2}^{2}(t)\text{$\textsf{\textbf{)%
}}$}/4\text{\textbf{,}} & t>\tau ,%
\end{array}%
\right.
\end{equation}
which involves no constant total quantum correlations, as shown in Fig. 1(b).

This phenomenon is quite different from the cases measured by $\mathcal{D}$
or $Q_{\mathcal{R}}$. Although they share a common critical point $\tau$,
the original `decoherence-free' area of total quantum correlations discovered by $
\mathcal{D}$ or $Q_{\mathcal{R}}$ \cite{SCD-exp,SCDcons1,SCDcons2} does not
appear in the measure of $Q_{\mathcal{S}}$.

%%%%%%%%%%%%%%%%%%%%%%%%%%%%%%%%%%%%%%%%
\begin{figure}[tbp]
\centering
\includegraphics[width=3.5in]{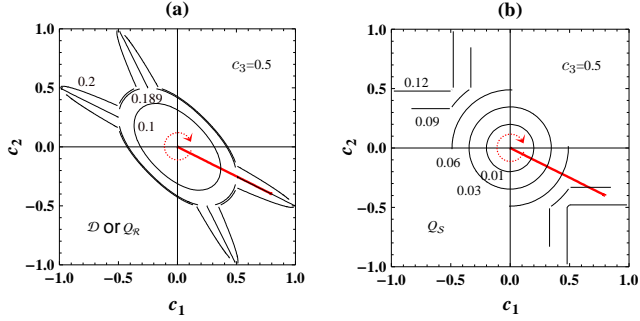}
\caption{Contour maps for total quantum correlations of
Bell-diagonal states using the definitions of (a) $\mathcal{D}$ or  $Q_{\mathcal{R}}$,
and (b) $Q_{\mathcal{S}}$, respectively. The red straight lines [$
c_{2}(t)=\protect\epsilon c_{1}(t)$ with $\protect\epsilon=c_{2}(0)/c_{1}(0)$]
represent the trajectories of the Bell-diagonal state
under non-Markovian dephasing channels with the initial conditions \textsf{%
\textbf{(}}$c_{1}(0),c_{2}(0),c_{3}(0)$\textsf{\textbf{)}}$\mathbf{=(}%
0.8,-0.4,0.5\textbf{)}$. The red-dotted circular arrows represent the
possible distribution of other lines through the origin of coordinate,
corresponding to trajectories of Bell-diagonal states with other possible
initial conditions.}
\end{figure}
%%%%%%%%%%%%%%%%%%%%%%%%%%%%%%%%%%%%%%%%

To be more clarified, we have plotted in Fig. 2 the contour maps of quantum
discord in a two-dimensional coordinate space with $c_3=0.5$. Recalling $%
c_{2}(t)=\epsilon c_{1}(t)$ with $\epsilon $= $\frac{c_{2}(0)}{c_{1}(0)}$,
the possible trajectories under the non-Markovian dephasing channels should
be the straight lines crossing the origin of coordinate (the red-dotted
circular arrows represent the distribution of line's slope, i.e., different
initial conditions). For a special case of $c_{1}(0)=0.8$ and $%
c_{2}(0)=-0.4, $ the trajectory of the Bell-diagonal states under
decoherence is depicted as the red lines in Fig. 2. Clearly, the red
line coincides with the straight contour line in Fig. 2(a), which
means the quantum discord will not be spoiled in this
`decoherence-free' area \cite {SCD-exp,SCDcons1,SCDcons2}. In
addition, there are three other `decoherence-free' areas (See the
black straight contour lines, one for $\epsilon=-c_3$ and the other
two for $\epsilon=-1/c_3$\cite{Note}) as depicted in Fig. 2(a). For other values of $\epsilon$ (initial conditions), the red line will always cross the contour lines and no constant quantum discord will take place.
However, in Fig. 2(b), the red straight lines always definitely go through
the contour lines, which is a direct illustration of no constant
quantum discord by $Q_{\mathcal{S}}$.

Since all the three quantities are to measure the total quantum
correlations, we wonder why $Q_{\mathcal{S}}$ is incompatible with
$\mathcal{D}$ and $Q_{\mathcal{R}}$ in describing the dynamics of
Bell-diagonal states under non-Markovian dephasing channels? As both
$Q_{\mathcal{R}}$ and $Q_{\mathcal{S}}$ are defined from geometric
perspective, we guess the imcompatibility is from the fact that the
nearest zero-discord state, belonging to an arbitrary Bell-diagonal
state and measured by the square of Hilbert-Schmidt norm, is
different from the closest classical state quantified by the
relative entropy.

%%%%%%%%%%%%%%%%%%%%%%%%%%%%%%%%%%%%%%%%
\begin{figure}[tbp]
\centering
\includegraphics[width=2.5in,bb=79pt 443pt 379pt 772pt]{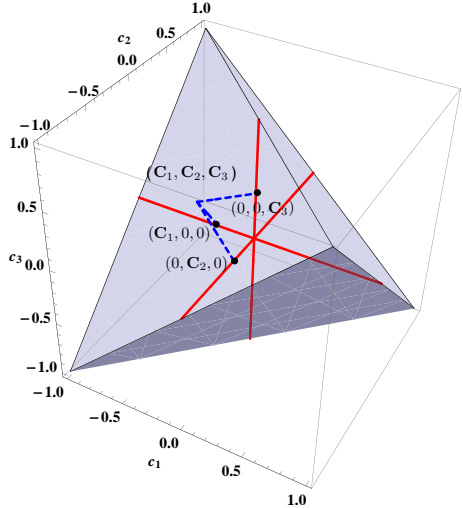}
\caption{The set of Bell-diagonal states with three parameters
\textbf{(}$c_{1},c_{2},c_{3}$\textbf{).} The red lines represent
zero-discord states. The blue-dashed lines connect to the possible
nearest zero-discord states for a given Bell-diagonal state
($\mathbf{C}_{1},\mathbf{C}_{2},\mathbf{C}_{3}$).} \label{fig3}
\end{figure}
%%%%%%%%%%%%%%%%%%%%%%%%%%%%%%%%%%%%%%%%

The guess is checked below. As demonstrated in Ref.
\cite{discord-G}, the set of zero-discord in a three-dimensional
space spanned by ($c_{1},c_{2},c_{3}$) includes three mutually
perpendicular lines \{$c_{j}\in \lbrack -1,1]$ $|$ $c_{k}=0,k\neq j$
$(j=1,2,3)$\} (red lines in Fig. 3) and the zero-discord states can
be written as $\Omega=(1+c_{j}\sigma _{j}\otimes \sigma _{j})/4$
($j=1,2,3$). For an arbitrarily given Bell-diagonal state $\rho$ denoted by
($\mathbf{C}_{1},\mathbf{C}_{2},\mathbf{C }_{3}$), the distance to
the zero-discord states measured by the square of Hilbert-Schmidt
norm can be calculated as
\begin{align}
||\rho-\Omega||^{2}  & =Tr\{\textsf{\textbf{(}}\frac{(\mathbf{C}_{j}%
-c_{j})\sigma_{j}\otimes\sigma_{j}+\mathbf{C}_{k}\sigma_{k}\otimes\sigma
_{k}+\mathbf{C}_{l}\sigma_{l}\otimes\sigma_{l}}{4}\textsf{\textbf{)}}%
^{2}\}\nonumber\\
& =\frac{(\mathbf{C}_{j}-c_{j})^{2}+\mathbf{C}_{k}^{2}+\mathbf{C}_{l}^{2}}%
{4}(j\neq k\neq l).
\end{align}
Clearly, $||\rho -\Omega ||^{2}$ reaches the minimum
only when $\mathbf{C} _{j}-c_{j}=0.$ Therefore, the possible nearest
zero-discord states should be from ($\mathbf{C}_{1},0,0$),
($0,\mathbf{C}_{2},0$), and ($0,0,\mathbf{C}_{3} $), dependent on
the magnitude among $\left\vert \mathbf{C}_{1}\right\vert$ ,
$\left\vert \mathbf{C}_{2}\right\vert ,$ and $\left\vert \mathbf{C}
_{3}\right\vert $. For example, when
$\mathbf{C}_{1}>\mathbf{C}_{2}>\mathbf{C }_{3}>0,$ the nearest
zero-discord state is ($\mathbf{C}_{1},0,0$).

On the other hand, as $\lambda _{00}$ and $\lambda _{01}$ are larger
than $ \lambda _{10}$ and $\lambda _{11}$ in the case of
$\mathbf{C}_{1}>\mathbf{C} _{2}>\mathbf{C}_{3}>0$ (in other cases
with arbitrary ordinal relation for $ \left\vert
\mathbf{C}_{1}\right\vert ,\left\vert \mathbf{C}_{2}\right\vert ,$
and $\left\vert \mathbf{C}_{3}\right\vert $, the proof is similar),
we have $ \Lambda =\lambda _{00}+\lambda
_{01}=(1+\mathbf{C}_{1})/2$. Recalling the requirements for the
closest classical state measured by the relative entropy
\cite{classical}, we can obtain $\lambda _{00}=\lambda _{01}=\Lambda
/2$ and
$\lambda _{10}=\lambda _{11}=(1-\Lambda )/2$, i.e., $\mathbf{C}_{2}=\mathbf{C%
}_{3}=0.$ Therefore, the closest classical state measured by the relative
entropy is also ($\mathbf{C}_{1},0,0$), which coincides with the nearest
zero-discord state measured by the square of Hilbert-Schmidt norm.

Since for an arbitrary Bell-diagonal state, the nearest zero-discord
state measured by $Q_{\mathcal{S}}$ is just the closest classical
state quantified by $Q_{\mathcal{R}},$ the discrepancy we discovered
in this work must be resulted from the intrinsic nature of the
square of Hilbert-Schmidt norm and the relative entropy.

To summarize, we have investigated quantum discord of Bell-diagonal
states under decoherence by three different definitions. The differences and
similarities by using the three measures have been presented and discussed.
The study of critical point under non-Markovian environment might be helpful
for prolonging total quantum correlations under decoherence.

Finally, it would be really interesting to further explore the
`decoherence-free' area of total quantum correlations, which appears in
the originally defined quantum discord $\mathcal{D}$ and relative
entropy based $Q_{\mathcal{R}}$, but disappears from the geometric
perspective $Q_{\mathcal{S}}$. It may lead to a more fundamental
quantum information problem, that is, which measure, $\mathcal{D}$
($Q_{\mathcal{R}}$) or $Q_{\mathcal{S}}$, is more accurate to
characterize total quantum correlations?

This work is supported by the National Natural Science Foundation of China
under Grant No. 10774163.


\begin{thebibliography}{99}
\bibitem{entanglement} R. Horodecki \textit{et al}., Rev. Mod. Phys. \textbf{%
81}, 865 (2009).

\bibitem{book QIC} M. A. Nielsen and I. L. Chuang, \textit{Quantum
Computation and Quantum Information} (Cambridge University Press, Cambridge,
England, 2000).

\bibitem{DQC1} E. Knill and R. Laflamme, Phys. Rev. Lett. \textbf{81}, 5672
(1998); A. Datta \textit{et al}., \textit{ibid}. \textbf{100}, 050502
(2008); B. P. Lanyon \textit{et al}., \textit{ibid}. \textbf{101}, 200501
(2008); A. Datta and S. Gharibian, Phys. Rev. A \textbf{79}, 042325 (2009).

\bibitem{discord} H. Ollivier and W. H. Zurek, Phys. Rev. Lett. \textbf{88},
017901 (2001).

\bibitem{discord-dengjia} L. Henderson and V. Vedral, J. Phy. A \textbf{34},
6899 (2001).

\bibitem{add1} J. Oppenheim \textit{et al.}, Phys. Rev. Lett. \textbf{89},
180402 (2002).

\bibitem{D-Maxwell} W. H. Zurek, Phys. Rev. A \textbf{67}, 012320 (2003); A.
Brodutch and D. R. Terno, \textit{ibid}. \textbf{81}, 062103 (2010).

\bibitem{add2} B. Groisman \textit{et al.}, Phys. Rev. A \textbf{72}, 032317
(2005).

\bibitem{D-2qubit} S. Luo, Phys. Rev. A \textbf{77}, 042303 (2008).

\bibitem{D-XX} R. Dillenschneider, Phys. Rev. B \textbf{78}, 224413 (2008);
M. S. Sarandy, Phys. Rev. A \textbf{80}, 022108 (2009); Y. X. Chen and S. W.
Li, \textit{ibid}. \textbf{81}, 032120 (2010); T. Werlang and G. Rigolin,
\textit{ibid}. \textbf{81}, 044101 (2010); J. Maziero \textit{et al}.,
\textit{ibid}. \textbf{82}, 012106 (2010); T. Werlang \textit{et al}., Phys.
Rev. Lett. 105, 095702 (2010); Z.-Y. Sun \textit{et al}., Phys. Rev. A
\textbf{82}, 032310 (2010).

\bibitem{NLB} M. Piani \textit{et al}., Phys. Rev. Lett. \textbf{100},
090502 (2008); S. Luo and W. Sun, Phys. Rev. A \textbf{82}, 012338 (2010).

\bibitem{D0} A. Shabani and D. A. Lidar, Phys. Rev. Lett. \textbf{102},
100402 (2009); A. Ferraro \textit{et al}., Phys. Rev. A \textbf{81}, 052318
(2010).

\bibitem{D-fields} A. Datta, Phys. Rev. A \textbf{80}, 052304 (2009); J.
Wang \textit{et al}., \textit{ibid}. \textbf{81}, 052120 (2010).

\bibitem{Hundun} Y. Y. Xu \textit{et al}., Europhys. Lett. \textbf{92}, 10005 (2010).

\bibitem{D-X} M. Ali \textit{et al}., Phys. Rev. A \textbf{81}, 042105
(2010).

\bibitem{D-NMR} D. O. Soares-Pinto \textit{et al}., Phys. Rev. A \textbf{81}%
, 062118 (2010).

\bibitem{D-CV} P. Giorda and M. G. A. Paris, Phys. Rev. Lett. \textbf{105},
020503 (2010); G. Adesso and A. Datta, \textit{ibid}. \textbf{105}, 030501
(2010); R. Vasile \textit{et al}., Phys. Rev. A \textbf{82}, 012313 (2010).

\bibitem{MD} T. Werlang \textit{et al}., Phys. Rev. A \textbf{80}, 024103
(2009).

\bibitem{SCDchan} J. Maziero \textit{et al}., Phys. Rev. A \textbf{80},
044102 (2009).

\bibitem{SCD-exp} J.-S. Xu \textit{et al}., Nature Commun. \textbf{1}, 7
(2010).

\bibitem{SCDcons1} L. Mazzola \textit{et al}., Phys. Rev. Lett. \textbf{104}%
, 200401 (2010).

\bibitem{SCDcons2} M. D. Lang and C. M. Caves, Phys. Rev. Lett. \textbf{105}%
, 150501 (2010).

\bibitem{SCDchan2} X.-M. Lu \textit{et al}., Quantum Inf. Comput. \textbf{10}%
, 0994 (2010).

\bibitem{nonMD} B. Wang \textit{et al}., Phys. Rev. A \textbf{81}, 014101
(2010); F. F. Fanchini \textit{et al}., \textit{ibid}. \textbf{81}, 052107
(2010).

\bibitem{discord-R} K. Modi \textit{et al}., Phys. Rev. Lett. \textbf{104},
080501 (2010).

\bibitem{discord-G} B. Daki\'{c} \textit{et al}., Phys. Rev. Lett. \textbf{%
105}, 190502 (2010).

\bibitem{discord-GL} S. Luo and S. Fu, Phys. Rev. A \textbf{82}, 034302
(2010).

\bibitem{Note0} For bipartite quantum systems as an example, the classical state is of the form
$\rho_{AB}=\sum_{ij}p_{ij}\Pi_{i}^{A}\otimes\Pi_{j}^{B},$ where
\{$p_{ij}$\} is some probability distribution and $\left\{
\Pi_{i}^{A}\right\}$ ($\left\{  \Pi_{j}^{B}\right\}  $) is the
eigenprojectors of $\rho_{A} =tr_{B}\rho_{AB}$ ($\rho_{B}=
tr_{A}\rho_{AB}$). On the other hand, the semiquantum state
(zero-discord state) is of the form $\rho_{AB}=\sum_{i}
p_{i}\Pi_{i}^{A}\otimes\rho_{i}^{B}.$ For more details about the
classical and semiquantum states please see in Phys. Rev. A
\textbf{77}, 022301 (2008).

\bibitem{simianti} R. Horodecki and M. Horodecki, Phys. Rev. A \textbf{54},
1838 (1996).

\bibitem{nonM} T. Yu and J. H. Eberly, Opt. Commun. \textbf{283}, 676 (2010).

\bibitem{Note}
Provided the initial conditions $|c_{2}(0)|\geq
|c_{1}(0)|,|c_3|$ and $\epsilon =-1/c_3$, we will have Eqs. (7)$\sim $(9) and (12) with $c_{1}(t)$ [$c_{2}(t)$] replaced by $c_{2}(t)$ [$c_{1}(t)$].

\bibitem{classical} V. Vedral and M. B. Plenio, Phys. Rev. A \textbf{57},
1619 (1998).



\end{thebibliography}
\end{document}